\documentclass[letterpaper, 10 pt, conference]{ieeeconf}
\IEEEoverridecommandlockouts                              
\usepackage[noadjust]{cite}
\usepackage{graphicx}
\usepackage{amsmath}
\usepackage{amssymb}
\usepackage{array}
\usepackage{stfloats}
\usepackage{floatrow}
\usepackage{tabularx}
\usepackage{multirow}
\usepackage{hhline}
\usepackage[export]{adjustbox}
\usepackage{pgfplots}
\usepackage[T1]{fontenc}
\usepackage[latin9]{inputenc}
\usepackage{epsfig}
\usepackage{collcell}
\usepackage{pgf}
\pgfplotsset{compat=1.14}
\usepackage{graphicx}
\usepackage{tabularx}
\newcolumntype{C}[1]{>{\centering\arraybackslash}p{#1}}
\usepackage{caption}
\usepackage{graphicx}
\usepackage{subcaption}
\usepackage[outdir=./]{epstopdf}

\usepackage{tabularx}
\newcolumntype{C}[1]{>{\centering\arraybackslash}p{#1}}

\def\BibTeX{{\rm B\kern-.05em{\sc i\kern-.025em b}\kern-.08em
    T\kern-.1667em\lower.7ex\hbox{E}\kern-.125emX}}
\begin{document}

\title{Understanding Global Reaction to the Recent Outbreaks of COVID-19: Insights from Instagram Data Analysis
\\
}

	
\author{Abdul Muntakim Rafi$^{1}$, Shivang Rana$^{2}$, Rajwinder Kaur$^{3}$, Q.M. Jonathan Wu$^{4}$, and Pooya Moradian Zadeh$^{5}$\\
	$^{1, 2, 4}$Department of Electrical \& Computer Engineering, \\$^{3, 5}$ School of Computer Science\\
	University of Windsor, Ontario, Canada\\
	\{rafi11, rana11r, kaur19s, jwu, pooya\}@uwindsor.ca
\thanks{*The first 3 authors had equal contribution to the paper. The dataset is available at www.kaggle.com/muntakimrafi/global-reaction-to-covid19-instagram-data-analysis}
}


\maketitle
\thispagestyle{empty}
\pagestyle{empty}

\begin{abstract}

The coronavirus disease, also known as the COVID-19, is an ongoing pandemic of a severe acute respiratory syndrome. 
The pandemic has led to the cancellation of many religious, political, and cultural events around the world. A huge number of people have been stuck within their homes because of unprecedented lockdown measures taken globally. This paper examines the reaction of individuals to the virus outbreak-through the analytical lens of specific hashtags on the Instagram platform. The Instagram posts are analyzed in an attempt to surface commonalities in the way that individuals use visual social media when reacting to this crisis. After collecting the data, the posts containing the location data are selected. A portion of these data are chosen randomly and are categorized into five different categories. We perform several manual analyses to get insights into our collected dataset. Afterward, we use the ResNet-50 convolutional neural network for classifying the images associated with the posts, and attention-based LSTM networks for performing the caption classification. This paper discovers a range of emerging norms on social media in global crisis moments. The obtained results indicate that our proposed methodology can be used to automate the sentiment analysis of mass people using Instagram data.

\end{abstract}


\section{Introduction}

Advent of social media presents a promising opportunity for emotion analysis of the mass people. Different approaches have successfully analyzed social media data to detect several conditions \cite{moreno2011associations}. There is a  vast wealth of psychological data encoded in visual social media data, such as photographs posted to Instagram. In this paper, we explore the possibility of analyzing visual data from Instagram to understand the global reaction to the recent outbreak of the coronavirus. 

It is reasonable to prioritize research into Instagram analysis for this problem. Instagram members currently contribute almost 100 million new posts per day \cite{alhabash2017tale}, and Instagram's rate of new users joining has recently outpaced Twitter, YouTube, LinkedIn, and even Facebook \cite{chaffey2016global}. Users give more importance to the visual content while sharing posts on Instagram. The posts are accompanied by captions as well. Instagram, among all social media, is more of personal space for the users. People tend to be less formal in their reactions. The data shared here can often have hidden complex meanings, which makes the Instagram posts most suitable for our task. Instagram data has proved to be a useful source for emotion analysis of the mass people. Instagram tends to be a place where people share personal emotions \cite{leaver2018visualising}.

According to \cite{zhou2020pneumonia}, the origin of the outbreak has been identified as Wuhan. The virus rapidly spread all over the globe. 
As of May 2020, more than 4 million cases of COVID-19 is confirmed worldwide \cite{gisan}, \cite{nation}. World Health Organisation (WHO) has declared a Global Health emergency, as the number of this novel virus keeps ascending globally \cite{who}. Worldwide 212 countries or territories have confirmed cases of this novel virus \cite{gisan}. A lot of everyday activities around the world have been restricted except for necessity and health circumstances. Acknowledging the rapid spread of the novel virus, there is a considerable impact of this epidemic on social networks and media as well. 

Understanding public reactions to a crisis is the first step to solving the situation. The initial assessment includes how the crisis is affecting people's lives. A concern that disrupts the everyday lives of more people is a more severe problem. Therefore, successfully identifying human reactions to a situation can lead to a successful estimation of the seriousness of the problem. Also, the mental health of individuals may deteriorate during crisis moments. As Instagram is a platform where people share their emotions, it is one of the best places to analyze the mental health of mass groups of people. On average, 100 million photos are uploaded daily on Instagram \cite{statista}. Manual analysis of all the information shared on a social media platform is almost impossible. Recently, deep learning algorithms have performed magnificently at extracting features from visual data. Therefore, we try to explore the feasibility of deep learning algorithms in analyzing Instagram data in this paper.

The major contributions of our paper are as follows:

\begin{enumerate}
  \item We collect a dataset consisting of 11973 Instagram posts with location.
  \item We manually annotate 2534 posts for getting insight into our collected Instagram data. 
  \item We explore the performance of deep learning algorithms in performing the analysis without human intervention.
\end{enumerate}

The rest of the paper is organized as follows. Section II describes the related researches conducted on the literature. Section III explains the deep learning terminologies used throughout the paper. Sections IV explains our proposed methodology in details, and section V contains all of our experiments and and results. Finally, we present our conclusion at Section VI. 

\section{Literature Review}

Researchers have found that mental health is strongly associated with social activities \cite{castillo2007diagnostic}. Photographs posted to Instagram contain different dimensions that might be analyzed for psychological insight. The photographs can be analyzed to determine different characteristics, such as the presence of people in the photographs or videos, the setting in nature or indoors, the time of the day, etc. Some research have been conducted in extracting statistical properties from images for performing various types of analysis \cite{wild2010graphical}. Instagram metadata offers additional information like how many comments or likes the post received. Finally, platform activity measures of different areas can yield clues to Instagram user's mental state. These dynamics were briefly discussed in \cite{reece2017instagram}. In particular, Lup et al. \cite{lup2015instagram} and Andalibi et al. \cite{andalibi2015depression} proposed to use Instagram data for mental health analysis. Recently, some works in the literature started using multi-modal analysis. In \cite{gomez2018learning}, the authors showed that using both textual and visual analysis allowed them to distinguish between different types of users and their interests. 

Instagram posts have been used to find out the emerging norms in social network dynamics. In \cite{leaver2018visualising}, researchers have used specific hashtags to pull off data from Instagram and discovered the nature of the reaction of general people to birth and death. They have raised a crucial privacy concern by showing that people create the first digital footprint of their children even before they are born. They also showed that people share their emotions on Instagram. In \cite{sensis2017sensis}, the authors showed that Instagram plays a vital role in body dissatisfaction and women's drive for thinness. In their research, it has been shown how people's lives are affected by the posts they see on social media. 

Previously, the manual or statistical feature-based analysis was used to infer decisions about social media data. The advancement of deep learning has made it possible to learn from images and associated text in a self-supervised way. The idea originated from the works in \cite{norouzi2013zero}, where the researchers proposed that instead of predicting the ImageNet classes of the image, deep learning models can be used to infer the Word2Vec representations of their corresponding labels. Researchers used the idea to analyze social media data. In \cite{kuo2014discovering}, the authors proposed a method to mine data from different social media platforms and modalities related to New York City and examined the behavior of the citizens. In \cite{garcia2015identification}, the authors used geo-located social media photographs to identify the most visited places by tourists in major European cities. In \cite{chang2016instagram}, the authors used popular hashtags on Instagram to find out cultural differences between different cities and neighborhoods. In \cite{salvador2017learning}, the authors proposed a joint embedding of pictures of food and recipes. They were able to achieve interesting results on an image-recipe retrieval task. In \cite{boy2017reassembling}, the authors analyzed Instagram posts to study how the different neighborhoods, events, and cultures of the city are presented on Instagram. Researchers detected race, age, and gender of people from New York City Instagram images in \cite{singh2017towards}. They have also analyzed the social diversity of different neighborhoods and compare them to census-based metrics.

\section{Preliminaries}

\subsection{Convolutional Neural Network (CNN)} 

CNNs are a class of deep learning networks that can infer intricate visual patterns and symmetries from a set of images. It learns filters of specific kernel size and understands these patterns after passing the images through the filters \cite{gu2018recent}. The CNN network that is used in this project is ResNet-50 \cite{he2016deep}. 

Mathematically, an image is convolved at every layer in a CNN and features are made more accessible for dense layer at the final classification task. Suppose $x_{m}^{l}$ is the $m^{th}$ input feature in the $l^{th}$ layer then the $n^{th}$ output feature in that $y_{n}^{l}$ is calculated as:

\begin{equation*}
    y_{n}^{l}=\sum_{m}^{M^{l-1}} w_{n, m}^{l} * x_{m}^{l}+b_{n}^{l}
    \label{circleeq}
\end{equation*}

Here, where $M^{l-1}$ is the number of input maps, * denotes convolution operation, and $b_{n}^{l}$ is the bias of the $n^{th}$ output map in the $l^{th}$ layer.

\subsection{Recurrent Neural Network (RNN)} 

RNNs are a class of deep learning networks that consists of a series of memory cells, which allows input from the previous memory cells and can infer the context from time-series data, then subsequently generate output \cite{sherstinsky2018fundamentals}. 

For defining how RNN computes mathematically it's output, following equation is described: $$ \begin{array}{l}o^{t}=f\left(h^{t} ; \theta\right) \\ \\ h^{t}=g\left(h^{t-1}, x^{t} ; \theta\right) \end{array}  $$ 

where $o^{t}$ is the output of the RNN at time $t, x^{t}$ is the input to the RNN at time $t,$ and $h^{t}$ is the state of the hidden layer(s) at time $t$.

\subsection{Attention based LSTM networks}

Aspect based sentiment analysis requires a sort of mechanism which can have a weighted focus on variegated words in a particular set of sentences. This mechanism is called attention, which gives different weights to words in a sentence while considering various types of aspects of the sentence \cite{wang2016attention}. 

Taking $H$ be a matrix comprising of hidden vectors $\left[h_{1}, \ldots, h_{N}\right]$ that the LSTM has yielded, where $d$ is the dimensionality of hidden layers \& $N$ is the length of the particular input sentence. Let $v_{a}$ represents the embedding of aspect and $e_{N}$ be a identity vector. The attention mechanism will produce an attention weight vector $\alpha$ and a weighted hidden representation $r$. Thereby, it allows the attention mechanism to focus on a particular segment of the sentence while considering one specific type of aspect\cite{wang2016attention}.

\[
\begin{array}{l}
M=\tanh \left(\left[\begin{array}{c}
W_{h} H \\
W_{v} v_{a} \otimes e_{N}
\end{array}\right]\right)  \label{} \\
\\
\alpha=\operatorname{softmax}\left(w^{T} M\right) \label{} \\
\\
r=H \alpha^{T} \label{}
\end{array}
\]
where, $M, \alpha , r, W_{h}, W_{v}$ and $w$ are
projection parameters.

The final sentence representation is given by:
\[
h^{*}=\tanh \left(W_{p} r+W_{x} h_{N}\right)
\]
where, $h^{*}, W_{p}$ and $W_{x}$ are projection parameters to be learned during training. 
$h^{*}$ is defined as the feature representation of a sentence with given an input aspect. Then linear layer to convert sentence vector to $e$, followed by softmax layer to transform $e$ into a conditional probability is implemented.
\[
y=\operatorname{softmax}\left(W_{s} h^{*}+b_{s}\right)
\]
where $W_{s}$ and $b_{s}$ are the parameters for softmax layer.
For aspect-based sentiment analysis, the input aspect embedding is appended with each word input vector.

\subsection{Transfer Learning}

Learning from a pre-trained neural network can be applied to a different application. Usually, the initial layers of networks are kept frozen, and the final layers are trained again on the new data. This method enables transferring the learning from one application to another similar type of application without requiring much data.

\begin{figure}[h!]
  \includegraphics[width=8.5 cm, height = 8.5 cm]{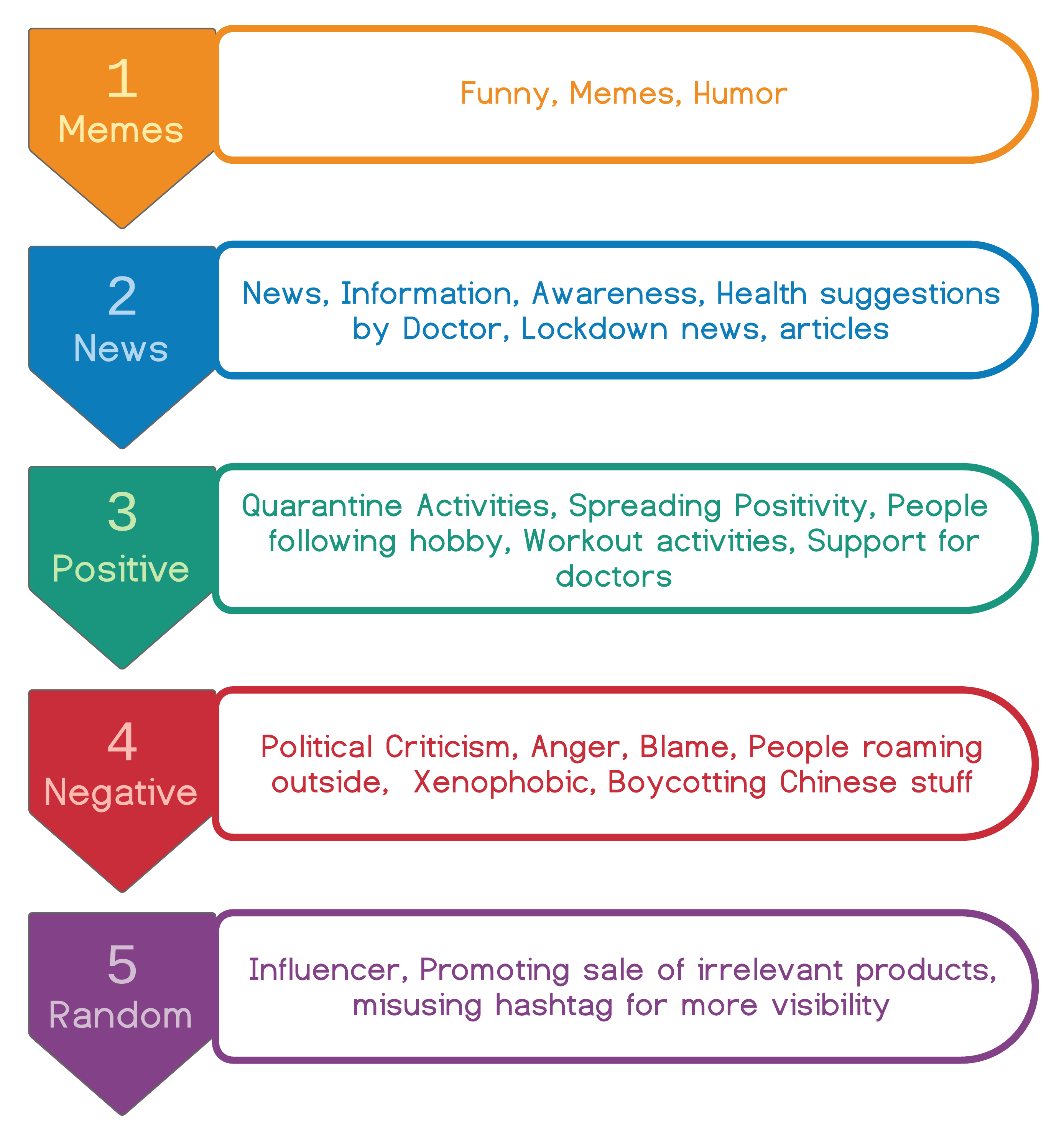}
  \caption{Different categories of data.}
  \label{fig:1}
\end{figure}

\section{Proposed Methodology}

Our proposed methodology can be defined by dividing the whole procedure into four tasks. We scrape data from Instagram, clean the data, annotate the data, and apply deep learning models on the collected data. We obtain results for understanding the sentiment from images and text captions both. 

\subsection{Data collection}

We use igramscrapper, a python API \cite{igramscraper} to collect data from Instagram using hashtags \#wuhan, \#corona, and \#coronavirus. We scrap the data from February 16 to March 20. We save the metadata related to each post in excel files and the corresponding media data (for videos we save the first frame). The metadata we store for each post are the post identifier, Instagram short code (i.e.,  B8o8MQHJbMc), created time, data type (image or video), source link to the post, low resolution image source link, high resolution image source link, post caption, owner id, number of likes received, number of comments received, and location name. All the posts we scrap from Instagram are public posts and visible to everyone.  

\subsection{Data Cleaning}

After gathering the data, we perform data cleaning. The data cleaning process is divided into several sub-tasks as following:

\begin{itemize}
    
    \item Data deduplication: In this step, we remove the data which are repeated.
    
    \item Data completeness: The part of data that is captured incompletely during the scraping process, can either be deleted or imputed to fill in meaning. In our case, missing data imputation may provide us wrong impressions about sentiments. For this reason, we delete the incomplete data points. 
    
    \item Corrupted Data: The corrupted data points are also removed.
    
\end{itemize}

\begin{figure}[h!]
  \includegraphics[width=8 cm, height = 5.5 cm]{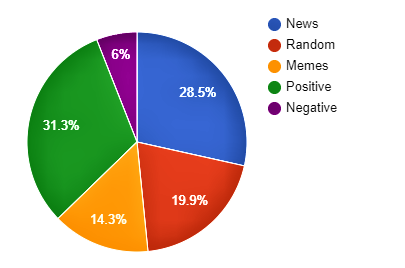}
  \caption{Class distribution of image annotations.}
  \label{fig:2_1}
\end{figure}

\begin{figure}[h!]
  \includegraphics[width=8 cm, height = 5.5 cm]{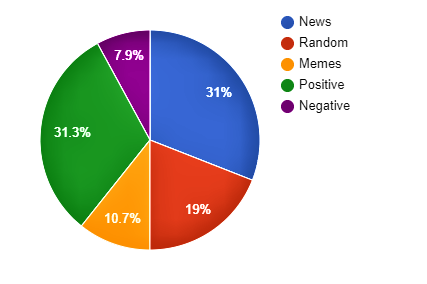}
  \caption{Class distribution of caption annotations.}
  \label{fig:2_2}
\end{figure}

\subsection{Data Annotation}

We categorize our annotated data into five categories, as shown in Fig. \ref{fig:1}. Humor, funny, and memes posts are classified as the first class. News articles, awareness posts, and information about cloud-19 and lockdown are marked as the second class. Happy, supportive, appreciable, positive, and nostalgic posts are put together in class 3. Negative posts regarding China, xenophobic, racist behavior, blaming, and protest against isolation are regarded as category 4. The class distribution of the image and caption annotations are illustrated in Fig. \ref{fig:2_1} and Fig. \ref{fig:2_2}. As depicted in the graphs, it can be inferred that the occurrence of "Positive" type and "Neutral/News" type annotations are the maximum , while "Negative" type annotation is observed the least.



\begin{figure*}[hbt!]
  \includegraphics[width=13 cm, height = 12 cm]{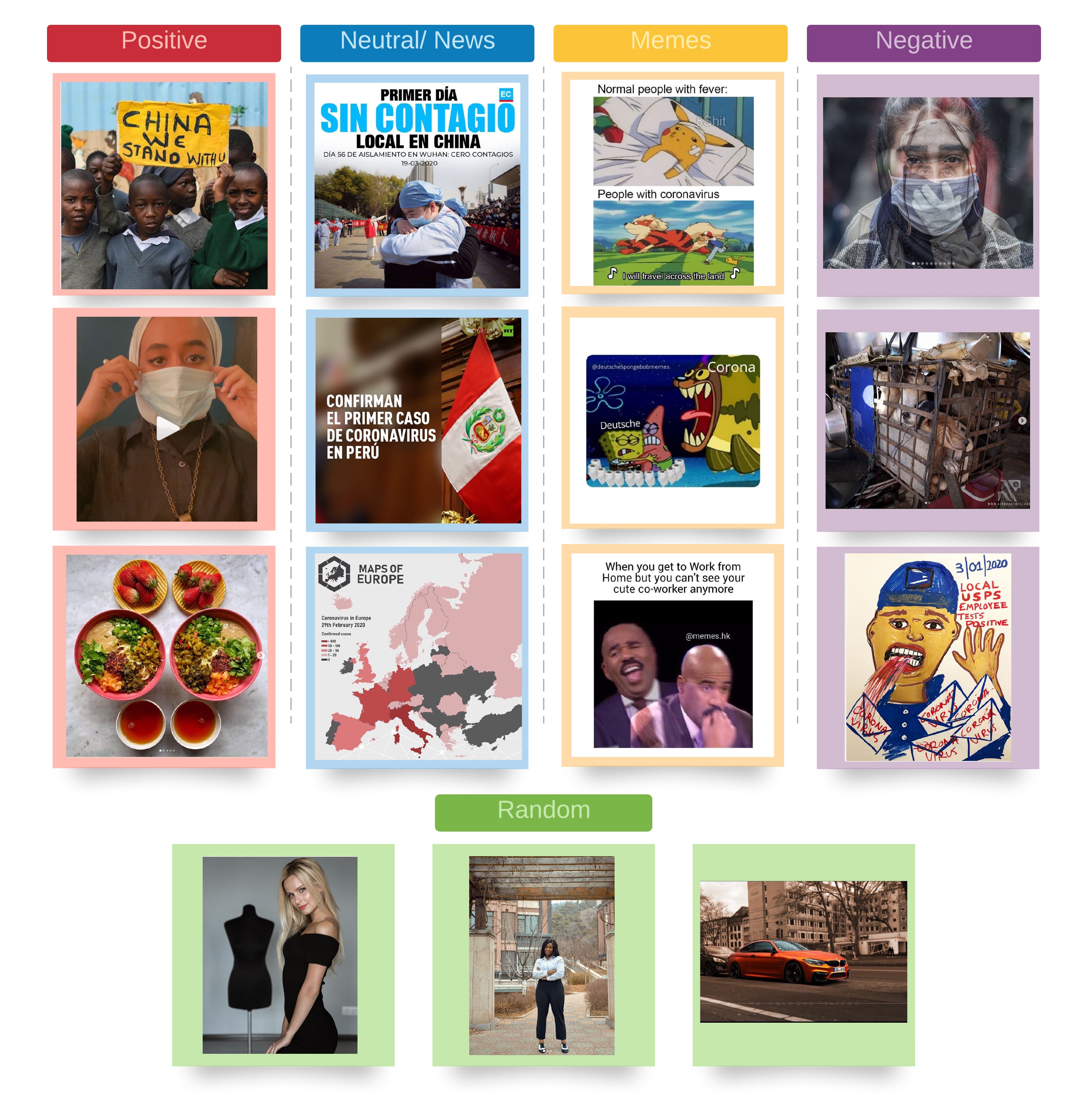}
  \caption{Samples from different classes.}
  \label{fig:13}
\end{figure*}

During the annotation process, we use Google Lens for translating the media content to help us with the annotation process. Google translate Chrome extension is used for translating the caption. Further discussion about which type of content is segregated in which kind of annotation is given below (see Fig. \ref{fig:13}).

\subsubsection{Memes/Humor Type}

This category comprises of any sort of meme or humor content, which mostly included people sharing memes surrounding toilet papers, corona beer, along with hilarious alternatives to sanitizers, and social distancing.

\subsubsection{News/Neutral Type}

This class covers posts, which were shared just to spread awareness about healthy habits like washing hands, wearing masks, and social distancing to be practiced during the pandemic. It also covers news information shared by a news channel on Instagram, along with advertisement about sanitizers, closure announcement of stores, info about toilet papers, masks, sanitizers available in a particular store. It also includes news reports about lower pollution and emptier streets. 

\subsubsection{Positive Type}

This category includes any Instagram post, which is targeted to spread not only physical but also mental positivity among the users. It covers posts regarding how people are spending their leisure time while quarantining themselves, which includes reading novels, dancing, exercising, playing video games, having movie watch parties, and singing. It also includes Instagrammers posting about the appreciation of the front-line workers, medical teams working during the pandemic, and wearing unique masks. This type of annotation also includes people posting their nostalgic past travel pictures and sharing pictures of how they are spending more time with their adorable pet dogs and cats. Medical doctors and nurses posting their selfies are also included in this category.

\subsubsection{Negative Type}
The negatively annotated data includes posts criticizing the political systems, xenophobic posts against China, anger posts against imposed lockdown, fear-mongering users by sharing content with people dying, ranting about loss of freedom, conducting public protests against the government. People blaming government and media for hiding factual data and spreading conspiracy theories also fall into this class.

\subsubsection{Random Type}

This class is used to separate Instagram influencers' posts and advertisements, which targets the promotion of unessential products by using the hashtags \#wuhan and \#coronavirus. It shows that the hashtags were misused to promote irrelevant agendas.

\subsection{Algorithm}

\begin{figure*}[hbt!]
  \includegraphics[width=14 cm, height = 8.5 cm]{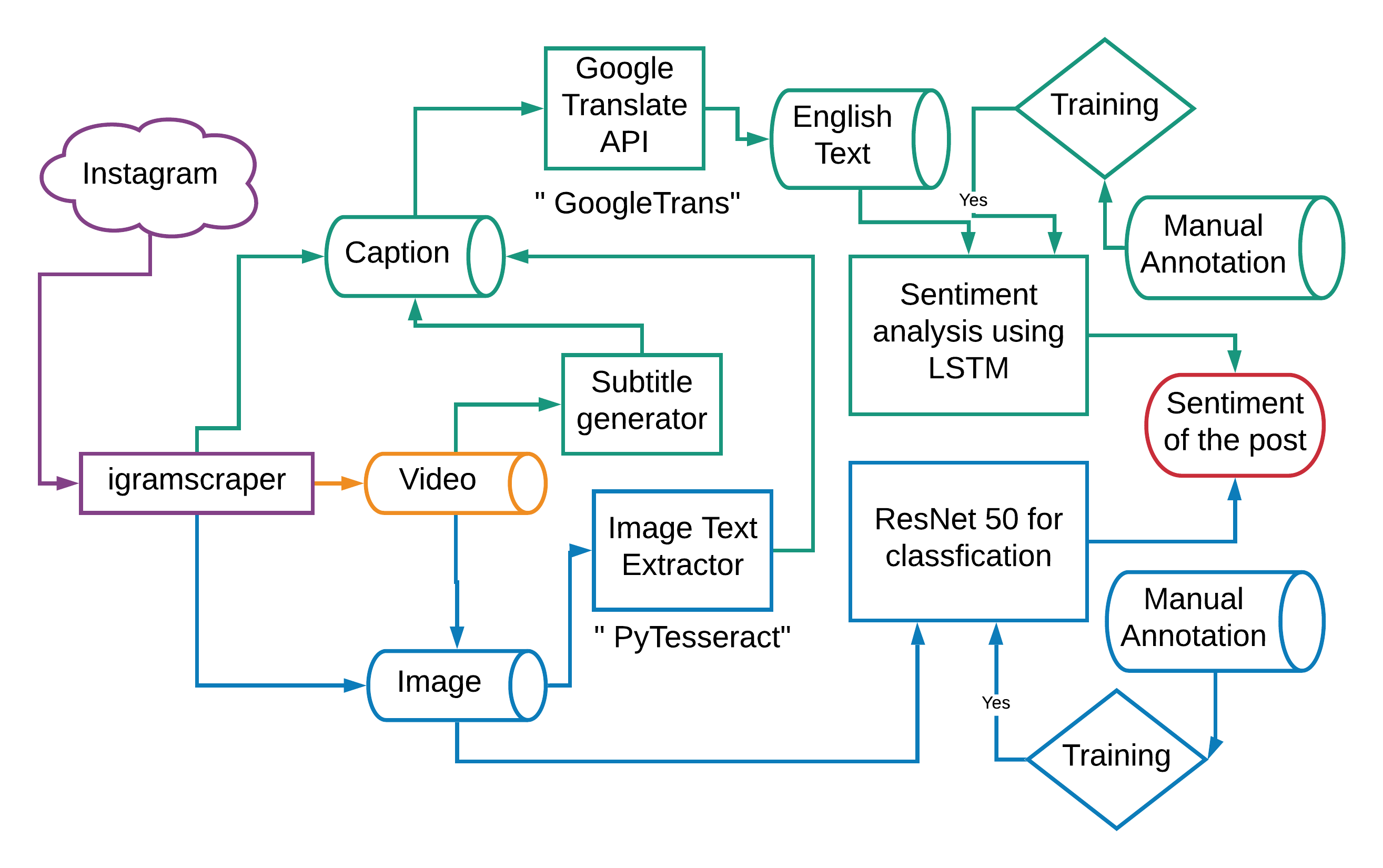}
  \caption{Proposed algorithm flowchart.}
  \label{fig:3}
\end{figure*}

We process the video data by taking the first frame of the video and save it as a single image which is stored along with the scraped image dataset. We extract audio from the video and run a subtitle generator on it. Thus, subtitles that are up-to 2 minutes are generated for the video and merged on top of stored caption data for the related post. Pytesseract API is used to extract text written on the image and saved along with the caption data stored for that post. Before deploying the deep learning models, first pre-processing steps are applied to caption text data and is translated into English using python API and trimmed up to word length of 300 words. Our proposed algorithms look at a single frame (image or the first frame of the video) for each post and the corresponding caption data.

The text data is used for training attention-based LSTM network for textual sentiment analysis. The image data is used for image classification using CNNs. We choose ResNet-50 for the image classification task. The proposed algorithm is illustrated in Fig. \ref{fig:3}.

\section{Experiments and results}

We perform several analyses to get insights into our collected dataset. We also explore the performance of deep learning algorithms for checking the feasibility of automatic analysis.  We use the Tensorflow library for implementing deep learning algorithms. All of the experiments regarding training and implementation of the models are performed in a hardware environment, which includes Intel Xeon, 2.60 GHz CPU and Nvidia Quadro M4000 (8 GB Memory) GPU.

\begin{figure}[hbt!]
  \includegraphics[width=8.5 cm, height = 5.5 cm]{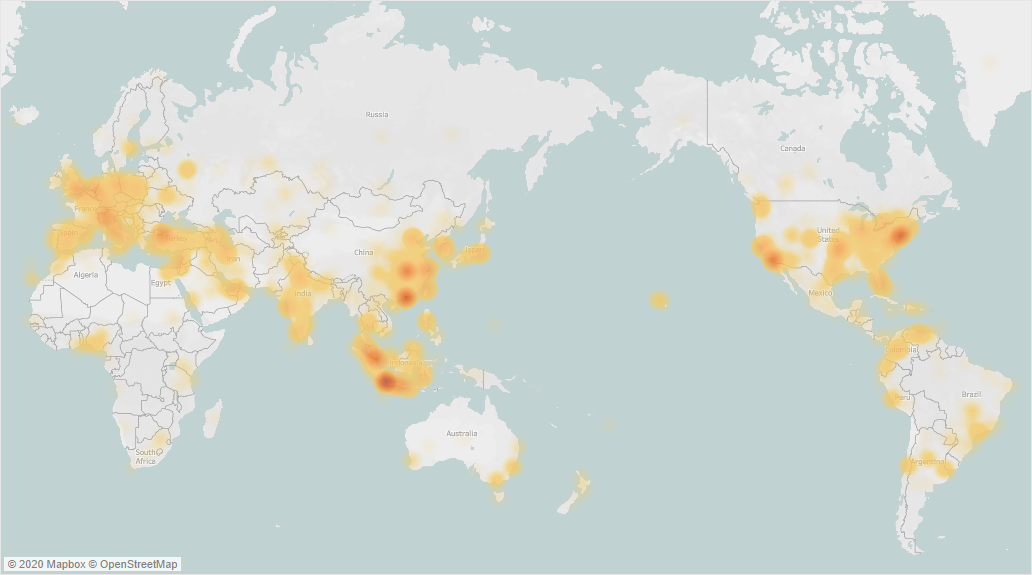}
  \caption{Geographic representation of Instagram activity for collected data based on number of posts.}
  \label{fig:4_1}
\end{figure}

\begin{figure}[hbt!]
  \includegraphics[width=8.5 cm, height = 5.5 cm]{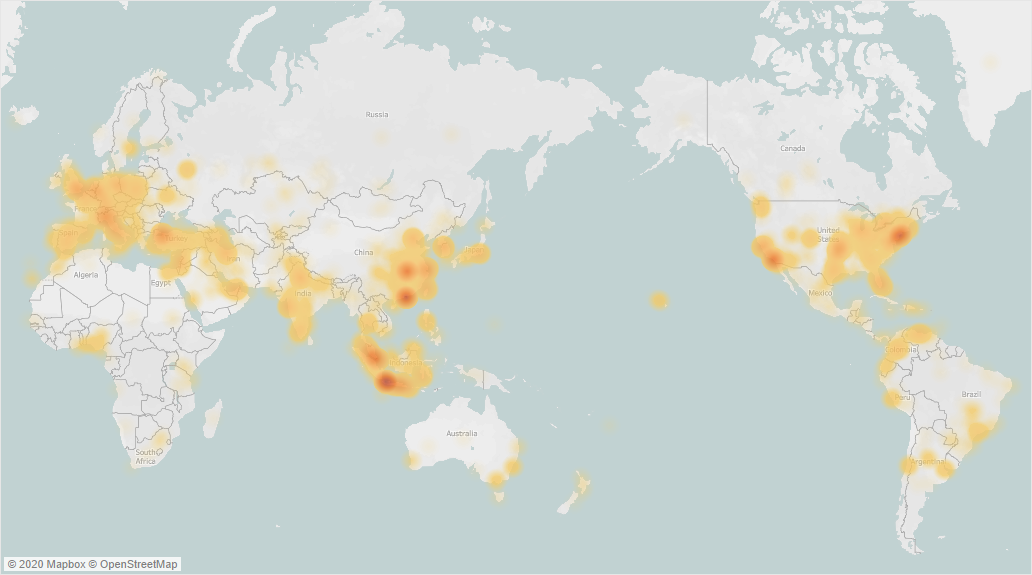}
  \caption{Geographic representation of Instagram activity for collected data based on the number of likes received.}
  \label{fig:4_2}
\end{figure}

\begin{figure}[hbt!]
  \includegraphics[width=8.5 cm, height = 5 cm]{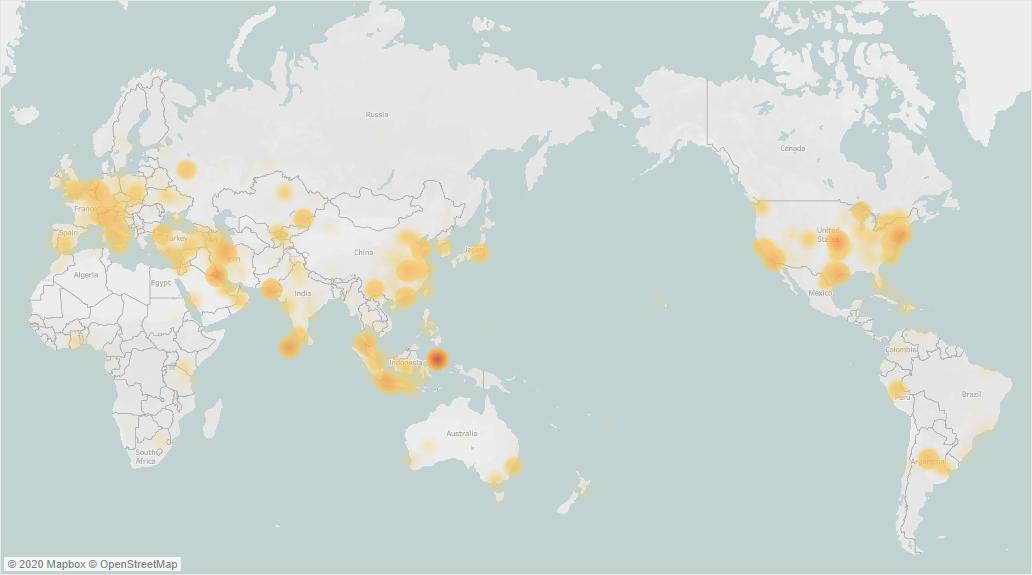}
  \caption{Geographic representation of Instagram activity\\ for collected data based on the number of comments received.}
  \label{fig:4_3}
\end{figure}

\begin{figure*}[hbt!]
  \includegraphics[width= 20 cm, height = 7 cm, center]{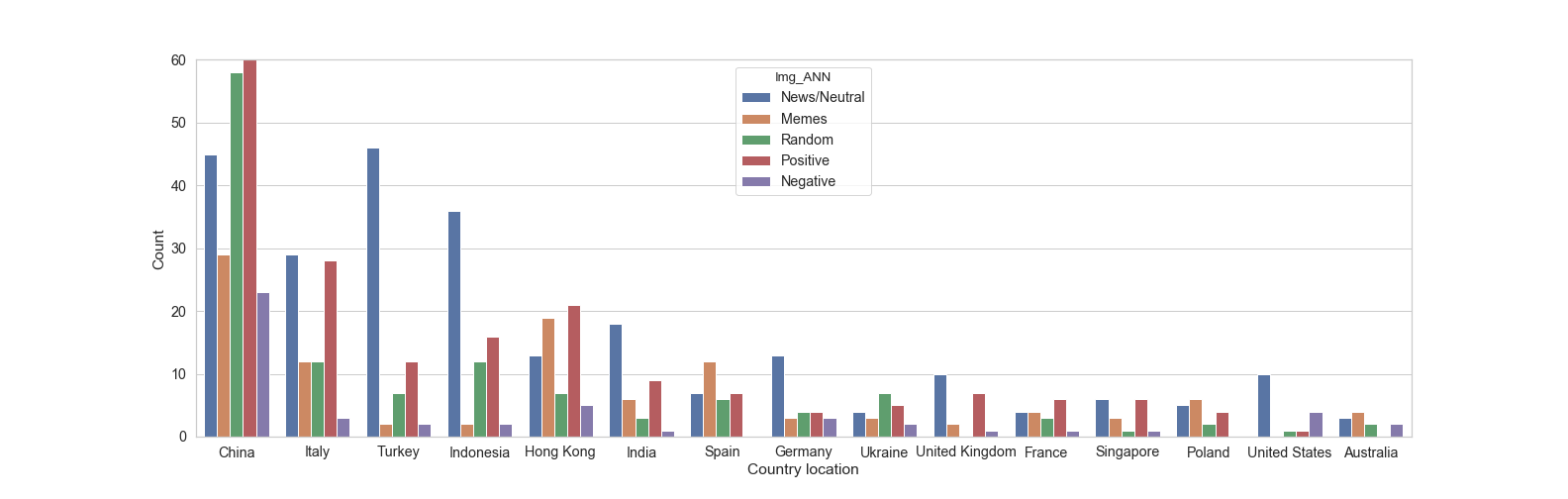}
  \caption{Bar Graph depicting sentiment in different countries (y-axis is capped at 60)}
  \label{fig:9}
\end{figure*}

\subsection{Location based analysis:}

We collect longitude and latitude for the location of the posts to perform several location-based analyses on our collected data. It enables us to see which regions have been most active and learn the degree of attention posts from different regions received from Instagram users. We use the number of likes and comments as indications for measuring attention. The data is visualized based on its longitude and latitude using Tableau. The geographic representation of data shows the intensity of global discussion worldwide on the Instagram platform. We illustrate the Instagram activity based on the number of posts, the number of likes received, and the number of comments received in Fig. \ref{fig:4_1}-\ref{fig:4_3}, respectively. In Fig. \ref{fig:4_1}, we can see that Indonesia, China, United States, Turkey, United Kingdom, Germany, Malaysia, and Italy have been most active concerning the number of posts posted on Instagram. It is also visible that Russia was quite inactive, given its 46M users \cite{statista}. We see somewhat similar trends in Fig. \ref{fig:4_2} and Fig. \ref{fig:4_3}.



\subsection{Country based analysis}

To understand the demographic viewpoint further, we extract the country name based on the location of the Instagram posts using the pycountry library. The top 15 countries with the maximum number of posts are illustrated with respect to to the count of the different class of annotations in Fig. \ref{fig:9}. Mostly, News/neutral, memes, and positive content are observed among the countries. The United States deviates from this trend as more negative posts are observed than positive posts. 

\subsection{Further insights into the data}

For getting further insights into our collected dataset, we plot several graphs by taking combinations of the following parameters: 

\begin{itemize}
    \item class of images in the Instagram posts (denoted by Img\_ANN),
    \item class of captions of the posts (denoted by Cap\_ANN),
    \item number of likes (denoted by likes\_count),
    \item number of comments (denoted by comments\_count).
\end{itemize}

The following Fig. \ref{fig:8} shows the class-overlap between the images and captions for the posts. It is visible that there are not many mismatches between the image annotation and the caption annotation for the Instagram posts.

\begin{figure}[!hbt]
  \centering
  \includegraphics[width=\columnwidth]{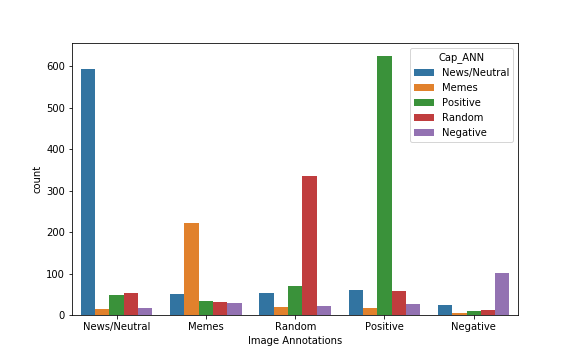}
  \caption{Class-overlap between the images and the captions for each post.}
  \label{fig:8}
\end{figure}





\begin{figure}[hbt!]
  \includegraphics[width= 8.5 cm, height = 6 cm]{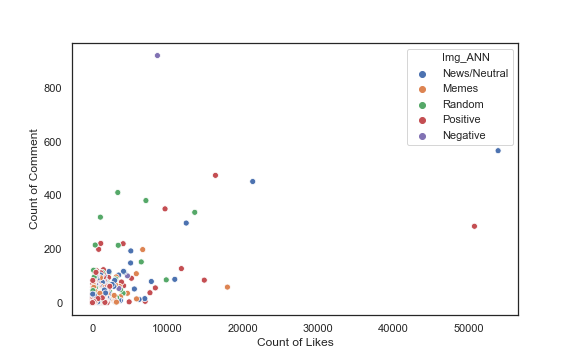}
  \caption{Number of comments vs. number of likes.}
  \label{fig:15}
\end{figure}

To understand the pattern between the number of likes received and the number of comments received, a scatter-point graph between these two parameters is plotted (see Fig. \ref{fig:15}). From the chart, we can infer two things. Firstly, negative posts tend to receive more comments than likes, as people provide their own opinion on that post without liking it. Secondly, positive posts receive more likes than comments. 

\begin{figure}[hbt!]
  \includegraphics[width= 8.5 cm, height = 6 cm]{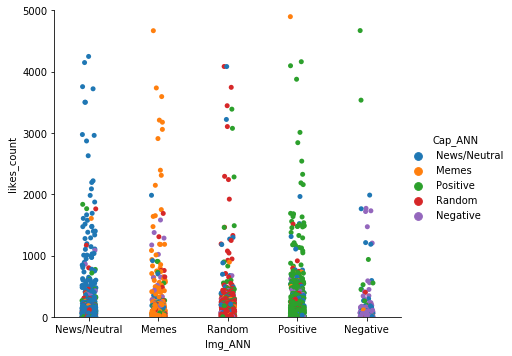}
  \caption{Scatterplot based on number of likes between image annotations and the corresponding caption annotations [y- scale is limited to 5000 likes]}
  \label{fig:14}
\end{figure}

A scatter-plot is shown with the following variables: the number of likes received, class of the image and the class of the and caption for each post (see \ref{fig:14}). We can see that posts with negative images but positive captions receive a high number of likes. It indicates that positive caption can change the class of the overall post and make it positive, despite containing a negative image. 

\begin{center}
	\begin{table}[]
		\centering
		\begin{tabular}{|C{0.1in}|C{1.5in}|C{0.9in}|C{0.3in}|}
			\hline
			\# & Model & Dataset & Acc.\\
			\hline
			1 & Textual Sentiment (Attention LSTM) \cite{wang2016attention}& Movie Review Dataset \cite{maas2011learning} & 87.7\%\\
			\hline
			2 & Textual Sentiment (Attention LSTM) \cite{wang2016attention}& \centering Our dataset & 73.6\% \\
			\hline
			3 & Image Sentiment Analysis (ResNet-50) \cite{gajarla2015emotion} & Reddit Memes Dataset \cite{goswami_2018} & 82.1\% \\
			\hline
			4 & Image Sentiment Analysis (ResNet-50) \cite{gajarla2015emotion} & Our Dataset & 79.2\% \\
			\hline
		\end{tabular}
		\caption{Performance of proposed deep learning algorithms.}
		\label{tab_result}
	\end{table}
\end{center}

\subsection{Performance of deep learning algorithms:}
The results we obtained from our experiments on the collected data are shown in Table \ref{tab_result}. We perform sentiment analysis on images and captions separately. For textual sentiment analysis; we use attention-LSTM \cite{wang2016attention} network and train it on Stanford's Movie Review Dataset \cite{maas2011learning}. We achieved 87.7\% accuracy. Later, we applied transfer learning and train the model on our annotated dataset. We achieved an accuracy of 73.6\%. 

For conducting sentiment analysis on images, we used ResNet-50 \cite{he2016deep} as suggested in the following paper \cite{gajarla2015emotion}. We used Reddit Memes Dataset \cite{goswami_2018} first to train the network and obtained 82.1 \% accuracy. Afterwards, we fine-tuned the pre-trained Resnet-50 on our dataset and achieved 79.2 \% accuracy. It should be noted that we exclude the `random' class while implementing our algorithms, and the networks are trained for classifying four classes.


\section{Conclusion}

In this paper, we have studied the reaction of Instagram users, in light of the recent outbreak of the coronavirus. It is essential to evaluate the public emotion in times of crisis and social media data, particularly Instagram, can be a valuable source for conducting such analysis in real-time. Consequently, we have collected in total, around 100k Instagram posts. Among them, around 14\% of the posts containing location data were used for our location-based analysis. We manually annotated around 20\% of the posts with location to perform our reaction analysis to the virus outbreak. The collected images and videos on Instagram reveal a range of emotions, highlighting social awareness, and mass mayhem. We found that users from most countries share positive and awareness posts. We also used deep learning algorithms to conduct the classification. Results show that the large-scale sentiment analysis can be automated with the help of deep learning algorithms. 

In our future work, we want to improve the accuracy of the deep learning models and perform the emotion analysis at a larger scale. 

\bibliographystyle{IEEEtran}
\bibliography{ref}

\begin{thebibliography}{10}
\providecommand{\url}[1]{#1}
\csname url@samestyle\endcsname
\providecommand{\newblock}{\relax}
\providecommand{\bibinfo}[2]{#2}
\providecommand{\BIBentrySTDinterwordspacing}{\spaceskip=0pt\relax}
\providecommand{\BIBentryALTinterwordstretchfactor}{4}
\providecommand{\BIBentryALTinterwordspacing}{\spaceskip=\fontdimen2\font plus
\BIBentryALTinterwordstretchfactor\fontdimen3\font minus
  \fontdimen4\font\relax}
\providecommand{\BIBforeignlanguage}[2]{{%
\expandafter\ifx\csname l@#1\endcsname\relax
\typeout{** WARNING: IEEEtran.bst: No hyphenation pattern has been}%
\typeout{** loaded for the language `#1'. Using the pattern for}%
\typeout{** the default language instead.}%
\else
\language=\csname l@#1\endcsname
\fi
#2}}
\providecommand{\BIBdecl}{\relax}
\BIBdecl

\bibitem{moreno2011associations}
M.~Moreno, D.~Christakis, K.~Egan, L.~Brockman, and T.~Becker, ``Associations
  between displayed alcohol references on facebook and problem drinking among
  college students: 125,'' \emph{Alcoholism: Clinical \& Experimental
  Research}, vol.~35, 2011.

\bibitem{alhabash2017tale}
S.~Alhabash and M.~Ma, ``A tale of four platforms: Motivations and uses of
  facebook, twitter, instagram, and snapchat among college students?''
  \emph{Social Media+ Society}, vol.~3, no.~1, p. 2056305117691544, 2017.

\bibitem{chaffey2016global}
D.~Chaffey, ``Global social media research summary 2016,'' \emph{Smart
  Insights: Social Media Marketing}, 2016.

\bibitem{leaver2018visualising}
T.~Leaver and T.~Highfield, ``Visualising the ends of identity: pre-birth and
  post-death on instagram,'' \emph{Information, Communication \& Society},
  vol.~21, no.~1, pp. 30--45, 2018.

\bibitem{zhou2020pneumonia}
P.~Zhou, X.-L. Yang, X.-G. Wang, B.~Hu, L.~Zhang, W.~Zhang, H.-R. Si, Y.~Zhu,
  B.~Li, C.-L. Huang \emph{et~al.}, ``A pneumonia outbreak associated with a
  new coronavirus of probable bat origin,'' \emph{nature}, vol. 579, no. 7798,
  pp. 270--273, 2020.

\bibitem{gisan}
``2019-ncov global cases study (by johns hopkins csse),''
  \url{https://gisanddata.maps.arcgis.com/apps/opsdashboard/index.html},
  accessed: 2020-01-31.

\bibitem{nation}
``Coronavirus live updates: Third case of coronavirus confirmed in london,
  ont., bringing canadian cases to four,''
  \url{https://nationalpost.com/news/world/coronavirus-live-updates-wuhan-virus-china-2019ncov},
  accessed: 2020-01-23.

\bibitem{who}
``Statement on the second meeting of the international health regulations
  (2005) emergency committee regarding the outbreak of novel coronavirus
  (2019-ncov),''
  \url{https://www.who.int/news-room/detail/30-01-2020-statement-on-the-second-meeting-of-the-international-health-regulations-(2005)-emergency-committee-regarding-the-outbreak-of-novel-coronavirus-(2019-ncov)},
  accessed: 2020-01-31.

\bibitem{statista}
``{statista},''
  \url{https://www.statista.com/statistics/578364/countries-with-most-instagram-users/}.

\bibitem{castillo2007diagnostic}
R.~Castillo, D.~Carlat, T.~Millon, C.~Millon, S.~Meagher, S.~Grossman, A.~P.
  Association \emph{et~al.}, ``Diagnostic and statistical manual of mental
  disorders,'' \emph{American Psychiatric Association Press, Washington, DC},
  2007.

\bibitem{wild2010graphical}
B.~Wild, M.~Eichler, H.-C. Friederich, M.~Hartmann, S.~Zipfel, and W.~Herzog,
  ``A graphical vector autoregressive modelling approach to the analysis of
  electronic diary data,'' \emph{BMC medical research methodology}, vol.~10,
  no.~1, p.~28, 2010.

\bibitem{reece2017instagram}
A.~G. Reece and C.~M. Danforth, ``Instagram photos reveal predictive markers of
  depression,'' \emph{EPJ Data Science}, vol.~6, no.~1, pp. 1--12, 2017.

\bibitem{lup2015instagram}
K.~Lup, L.~Trub, and L.~Rosenthal, ``Instagram\# instasad?: exploring
  associations among instagram use, depressive symptoms, negative social
  comparison, and strangers followed,'' \emph{Cyberpsychology, Behavior, and
  Social Networking}, vol.~18, no.~5, pp. 247--252, 2015.

\bibitem{andalibi2015depression}
N.~Andalibi, P.~Ozturk, and A.~Forte, ``Depression-related imagery on
  instagram. 2015 feb 28 presented at: The 18th acm conference companion on
  computer supported cooperative work \& social computing (pp.),'' 2015.

\bibitem{gomez2018learning}
R.~Gomez, L.~Gomez, J.~Gibert, and D.~Karatzas, ``Learning from barcelona
  instagram data what locals and tourists post about its neighbourhoods,'' in
  \emph{Proceedings of the European Conference on Computer Vision (ECCV)},
  2018, pp. 0--0.

\bibitem{sensis2017sensis}
Sensis, ``Sensis social media report 2017,'' 2017.

\bibitem{norouzi2013zero}
M.~Norouzi, T.~Mikolov, S.~Bengio, Y.~Singer, J.~Shlens, A.~Frome, G.~S.
  Corrado, and J.~Dean, ``Zero-shot learning by convex combination of semantic
  embeddings,'' \emph{arXiv preprint arXiv:1312.5650}, 2013.

\bibitem{kuo2014discovering}
Y.-H. Kuo, Y.-Y. Chen, B.-C. Chen, W.-Y. Lee, C.-C. Wu, C.-H. Lin, Y.-L. Hou,
  W.-F. Cheng, Y.-C. Tsai, C.-Y. Hung \emph{et~al.}, ``Discovering the city by
  mining diverse and multimodal data streams,'' in \emph{Proceedings of the
  22nd ACM international conference on Multimedia}, 2014, pp. 201--204.

\bibitem{garcia2015identification}
J.~C. Garc{\'\i}a-Palomares, J.~Guti{\'e}rrez, and C.~M{\'\i}nguez,
  ``Identification of tourist hot spots based on social networks: A comparative
  analysis of european metropolises using photo-sharing services and gis,''
  \emph{Applied Geography}, vol.~63, pp. 408--417, 2015.

\bibitem{chang2016instagram}
S.~Chang, ``Instagram post data analysis,'' \emph{arXiv preprint
  arXiv:1610.02445}, 2016.

\bibitem{salvador2017learning}
A.~Salvador, N.~Hynes, Y.~Aytar, J.~Marin, F.~Ofli, I.~Weber, and A.~Torralba,
  ``Learning cross-modal embeddings for cooking recipes and food images,'' in
  \emph{Proceedings of the IEEE conference on computer vision and pattern
  recognition}, 2017, pp. 3020--3028.

\bibitem{boy2017reassembling}
J.~D. Boy and J.~Uitermark, ``Reassembling the city through instagram,''
  \emph{Transactions of the Institute of British Geographers}, vol.~42, no.~4,
  pp. 612--624, 2017.

\bibitem{singh2017towards}
V.~K. Singh, S.~Hegde, and A.~Atrey, ``Towards measuring fine-grained diversity
  using social media photographs,'' in \emph{Eleventh International AAAI
  Conference on Web and Social Media}, 2017.

\bibitem{gu2018recent}
J.~Gu, Z.~Wang, J.~Kuen, L.~Ma, A.~Shahroudy, B.~Shuai, T.~Liu, X.~Wang,
  G.~Wang, J.~Cai \emph{et~al.}, ``Recent advances in convolutional neural
  networks,'' \emph{Pattern Recognition}, vol.~77, pp. 354--377, 2018.

\bibitem{he2016deep}
K.~He, X.~Zhang, S.~Ren, and J.~Sun, ``Deep residual learning for image
  recognition,'' in \emph{Proceedings of the IEEE conference on computer vision
  and pattern recognition}, 2016, pp. 770--778.

\bibitem{sherstinsky2018fundamentals}
A.~Sherstinsky, ``Fundamentals of recurrent neural network (rnn) and long
  short-term memory (lstm) network,'' \emph{arXiv preprint arXiv:1808.03314},
  2018.

\bibitem{wang2016attention}
Y.~Wang, M.~Huang, X.~Zhu, and L.~Zhao, ``Attention-based lstm for aspect-level
  sentiment classification,'' in \emph{Proceedings of the 2016 conference on
  empirical methods in natural language processing}, 2016, pp. 606--615.

\bibitem{igramscraper}
``{instagramscraper},'' \url{https://pypi.org/project/igramscraper/}.

\bibitem{maas2011learning}
A.~L. Maas, R.~E. Daly, P.~T. Pham, D.~Huang, A.~Y. Ng, and C.~Potts,
  ``Learning word vectors for sentiment analysis,'' in \emph{Proceedings of the
  49th annual meeting of the association for computational linguistics: Human
  language technologies-volume 1}.\hskip 1em plus 0.5em minus 0.4em\relax
  Association for Computational Linguistics, 2011, pp. 142--150.

\bibitem{gajarla2015emotion}
V.~Gajarla and A.~Gupta, ``Emotion detection and sentiment analysis of
  images.''

\bibitem{goswami_2018}
``{Reddit Memes Dataset},''
  \url{https://www.kaggle.com/sayangoswami/reddit-memes-dataset/version/1}.

\end{thebibliography}

\end{document}